\documentclass[conference]{IEEEtran}

\usepackage[colorinlistoftodos,prependcaption,textsize=tiny]{todonotes}
\usepackage{graphics}
\usepackage{courier}
\usepackage{cite}
\usepackage{color}
\usepackage{mathtools}
\usepackage{subfigure}
\usepackage{amssymb}
\usepackage{algorithm}
\usepackage{algpseudocode}
\usepackage{algorithmicx}
\usepackage{array}
\usepackage{color}

\usepackage[hyphens]{url}
\usepackage{paralist}
\usepackage[all]{xy}
\usepackage{tabu}
\usepackage{chngcntr}

\counterwithin{paragraph}{section}
\def\getsr{\stackrel{{\scriptscriptstyle\$}}{\leftarrow}}

 \IEEEoverridecommandlockouts

\begin{document}

\title{Multiparty Routing: Secure Routing for Mixnets}

\author{

\IEEEauthorblockN{Fatemeh Shirazi}

\IEEEauthorblockA{imec - COSIC KU Leuven}

\IEEEauthorblockA{Leuven, Belgium}

\and

\IEEEauthorblockN{Elena Andreeva}

\IEEEauthorblockA{imec - COSIC KU Leuven}

\IEEEauthorblockA{Leuven, Belgium}

\and

\IEEEauthorblockN{Markulf Kohlweiss}

\IEEEauthorblockA{Microsoft Research}

\IEEEauthorblockA{Cambridge, UK}

\and

\IEEEauthorblockN{Claudia Diaz}

\IEEEauthorblockA{imec - COSIC KU Leuven}

\IEEEauthorblockA{Leuven, Belgium}

}

\maketitle


\begin{abstract}

Anonymous communication networks are important building blocks for online privacy protection.
One approach to achieve anonymity is to relay messages through multiple routers, where each router shuffles messages independently.
To achieve anonymity, at least one router needs to be honest. In the presence of an adversary that is controlling a subset of the routers unbiased routing is important for guaranteeing anonymity. However, the routing strategy also influenced other factors such as the scalability and the performance of the system.
One solution is to use a fixed route for relaying all messages with many routers. If the route is not fixed the routing decision can either be made by the communication initiator or the intermediate routers.
However, the existing routing types each have limitations. For example, one faces scalability issues when increasing the throughput of systems with fixed routes.
Moreover, when the routing decision is left to the initiator, the initiator needs to maintain an up-to-date view of the system at all times, which also does not scale.
If the routing decision is left to intermediate routers the routing of the communication can be influenced by an adversary.
In this work, we propose a novel \emph{multiparty routing} approach for anonymous communication that addresses these shortcomings.
We distribute the routing decision and verify the correctness of routing to achieve routing integrity.
More concretely, we provide a mixnet design that uses our routing approach and that in addition, addresses load balancing.
We show that our system is secure against a global active adversary. 

\end{abstract} 




\section{Introduction}

Anonymous communication networks (ACNs) provide secure communication channels that hide not only data contents but also the communications' metadata, thus providing protection against traffic analysis. By routing data through multiple proxies, ACNs prevent network eavesdroppers from linking the source and destination of messages. Furthermore, ACNs enable senders to communicate with potentially malicious destinations without revealing their identity or location.

ACNs are typically overlay networks consisting of a set of dedicated routers, also called relays, that are connected via the Internet. The first ACN design in the early 1980s~\cite{Chaum:1981:mixnets} introduced the concept of \emph{mixes} for implementing anonymous email. These mixes are relays that collect a batch of $N$ equal-size messages, cryptographically transform them, and output them in a re-shuffled form, so that the output messages cannot be traced to their corresponding inputs based on message content, size, timing or order. The best known ACN is Tor~\cite{Dingledine:2004:Tor}. The Tor network is an ACN that consists of about seven thousands relays run by volunteers, and provides service to an estimated two million daily users who use it primarily to anonymously browse the web~\cite{Tor:2016:website}. Other examples of ACN networks include for example Mixminion~\cite{Danezis:2003:mixminion} an anonymous re-mailer, Freenet~\cite{Clarke:2001:FN,Clarke:2010:FN} used for anonymous file-sharing, and DC-Nets~\cite{Corrigan-Gibbs:2010:Dissent} that can be deployed for broadcast applications such as group messaging. 

The goal of ACNs is to anonymize communications by relaying them over multiple routers. There are three main types of anonymous routing in terms of how routers are chosen to form the path.

First, in \emph{deterministic routing}, the paths are predetermined by the system configuration. Chaum's original ACN proposal~\cite{Chaum:1981:mixnets} considered a sequence of mixes organized in a \emph{cascade}. Systems that adopted the cascade network topology in their designs include JAP \cite{Berthold:2000:JAP}, and voting systems, such as Helios~\cite{Adida:2008:Helios}. An advantage of cascades is that routing headers are unnecessary, since the next hop in the path is predetermined. Cascades however, are of limited practicality due to their poor scalability, limited anonymity, and lack of resilience to router failures~\cite{Kwon:2016:Atom}. 

Second, in \emph{source routing}, the full path is chosen by the initiator of the communication. This is the routing used by Tor~\cite{Dingledine:2004:Tor} and by anonymous remailers such as Mixmaster~\cite{moller:2003:mixmaster} and Mixminion~\cite{Danezis:2003:mixminion}. In comparison to cascades, source routing is more resilient against failures of routers~\cite{Shirazi:2015:MeasuringResilience}; it is more scalable, meaning that the network can grow to accommodate more users; and consequently, can provide better anonymity to those users. 

However, an important requirement in source routing is that the initiator must have a full view of the network, since otherwise partial network knowledge can be exploited by an adversary to fingerprint and identify users based on which relays they choose~\cite{Danezis:2008:RouteBridgingAttacks}. 
As a consequence, all users need to periodically download the full list of currently available routers, their addresses, public keys, and other routing-relevant information.
This is a major bottleneck for the scalability of the network, particularly if the routers have a high churn -- as is often the case in peer-to-peer networks.

Third, in \emph{hop-by-hop routing} each router locally decides on the next router of the path.  
Hop-by-hop routing is advantageous over source routing because it does not require entities to maintain an up-to-date view of the full network, as well as allows for better load balancing. 
The main disadvantage of hop-by-hop routing is that it is vulnerable to \emph{route capture} attacks, where a malicious relay can gain full control over the route of a message, so that it is relayed through adversarial relays~\cite{Danezis:2006:Fingerprinting}. 
The earliest ACN employing hop-by-hop routing is Crowds~\cite{Reiter:1998:Crowds}, and another example is Morphmix~\cite{Rennhard:2002:MorphMix}, which uses witnesses to mitigate route capture attacks. 

In this paper we propose \emph{multiparty routing}, a novel type of anonymous routing that broadens the design space with a fourth kind of routing. Our solution comes with important advantages over the previous routing approaches. Multiparty routing offers a scalability advantage over source routing: in our proposal the communication initiators need to only know one (or a small number) of routers, compared to the full view required in source routing. At the same time, we prevent the route capture attacks of hop-by-hop routing by decentralizing the routing decision using a multiparty computation technique. 
In addition to anonymity, our protocol guarantees the integrity of both shuffles and routing, which is novel for ACNs.

Our main contribution is a concrete multiparty routing protocol that employs cryptographic primitives, such as commitment schemes, signatures, and cryptographic hash functions. We combine our routing protocol with a mix network that employs provable shuffles, threshold decryption, and re-randomizable encryption to realize secure and verifiable anonymous communication, and load balancing. Our solution achieves security against global active adversaries and offers improved resilience and scalability against relay failures compared to deterministic and source routing. We achieve close to optimal load balancing by integrating relevant information about the throughput of relays into the routing strategy. 
Our protocol is a mid-latency anonymous communication network and thus can be deployed for applications that are more tolerant to latency such as anonymous wiki, micro-blogging, voting, and auctions. 

\section{Related Work}
\label{section:relatedwork}
Golle and Juels introduced Parallel Mixes~\cite{Golle:2004:ParallelMixes}, an
ACN that uses both deterministic and hop-by-hop routing in two different phases that they call 
\emph{rotation} and \emph{distribution} phases, respectively. 
Typical configurations start with a rotation phase, then go through a distribution phase, 
and finish with a rotation phase.
During a rotation phase, mixes exchange their entire output ciphertexts with another
mix in a rotating fashion (that is, the first mix passes its ciphertexts to the
second, and so on, and the last mix passes them to the first).
In the distribution phase, each mix splits output ciphertexts among other mixes. 
Our routing protocol has similarities with the distribution phase of Parallel
Mixes, but takes away one primary requirement of Parallel Mixes:  
Parallel Mixes require that in the distribution
phase mixes must not be aware of the sender of any of the ciphertexts. Consequently, users
have to submit input ciphertexts uniformly to the system and ciphertexts have to be mixed by at least
one honest mix during the rotation phase. This is necessary to prevent
route-capture attacks, because mixes in the
distribution phase are free to choose how to distribute their output ciphertexts to
other mixes (contrary to our protocol, where this decision is made jointly by multiple entities). 
However, these countermeasures are still insufficient when the adversary controls a fraction of the messages going through the network, as shown by an attack by Borisov~\cite{Borisov:2005:APM}. 

Movahedi et al.~\cite{Movahedi:2015:MPShuffling} proposed a multiparty shuffling protocol that allows multiple parties to jointly compute a private random permutation of input messages. The parties that perform the multiparty shuffling protocol are split up into quorums that use a sorting scheme on inputs that have been shared among the quorum to achieve random permutation. 
While the approach is resilient to parties aborting the protocol or dropping out because of technical difficulties, there is no mechanism for preventing traffic tampering before the shuffling starts. Therefore, parties who hold user inputs at the beginning of the shuffling can replace them with arbitrary adversarial inputs. 
The approach is therefore only suitable for passive adversary. Furthermore, this scheme requires that the links between honest parties must be private and cannot be observed by the adversary; and the adversary needs to be static, meaning it cannot corrupt alternative parties once the protocol is running.

\section{Goals, Assumptions, and System Architecture}\label{sec:protocol}
In this section, we describe the security model, the goals of the system, the entities and their roles, and the threat model.

\subsection{Adversary Model}
We assume a global adversary that can monitor all the communication links between entities in the system, including channels from users to the system and vice versa. 
The adversary is active, meaning that it can corrupt a subset of entities in the system, as well as submit messages herself.  
We assume that all cryptographic primitives are secure and cannot be broken by the adversary.  
Our protocol does not enforce restrictions on how many users need to be honest. 
However, a user is anonymous only among honest users that submitted messages to the network during the same time frame, and whose messages have traversed at least one honest relay.

\subsection{Goals of the System}
The goal of our system are as follows:
\begin{itemize}
 \item Anonymity: the adversary must not be able to link a message leaving the system to the user who sent it, provided that the message has traversed at least one honest relay. This goal is equivalent to sender anonymity that has been defined by Pfitzmann et al.\cite{Pfitzmann:2000:terminology}.
 \item Routing Integrity: the adversary must not be able to influence the route followed by a message, and consequently cannot bias the routing by forcing the data to only traverse adversarial relays. 
 \item Availability: the system must be robust to the adversarial removal of relays, and output all messages even if some relays are taken down once the routing has started. 
 \item Load-balancing: we aim at providing load balancing such that relays route an amount of traffic that is proportional to their throughput. 
\end{itemize}

\subsection{Entities and Roles}

We consider a broadcast channel that we call \emph{bulletin board}. Although the bulletin board is readable by all parties in the system: users and servers, users can not write into it.

\begin{figure}[h]
    \centering
\includegraphics[width=0.49\textwidth]{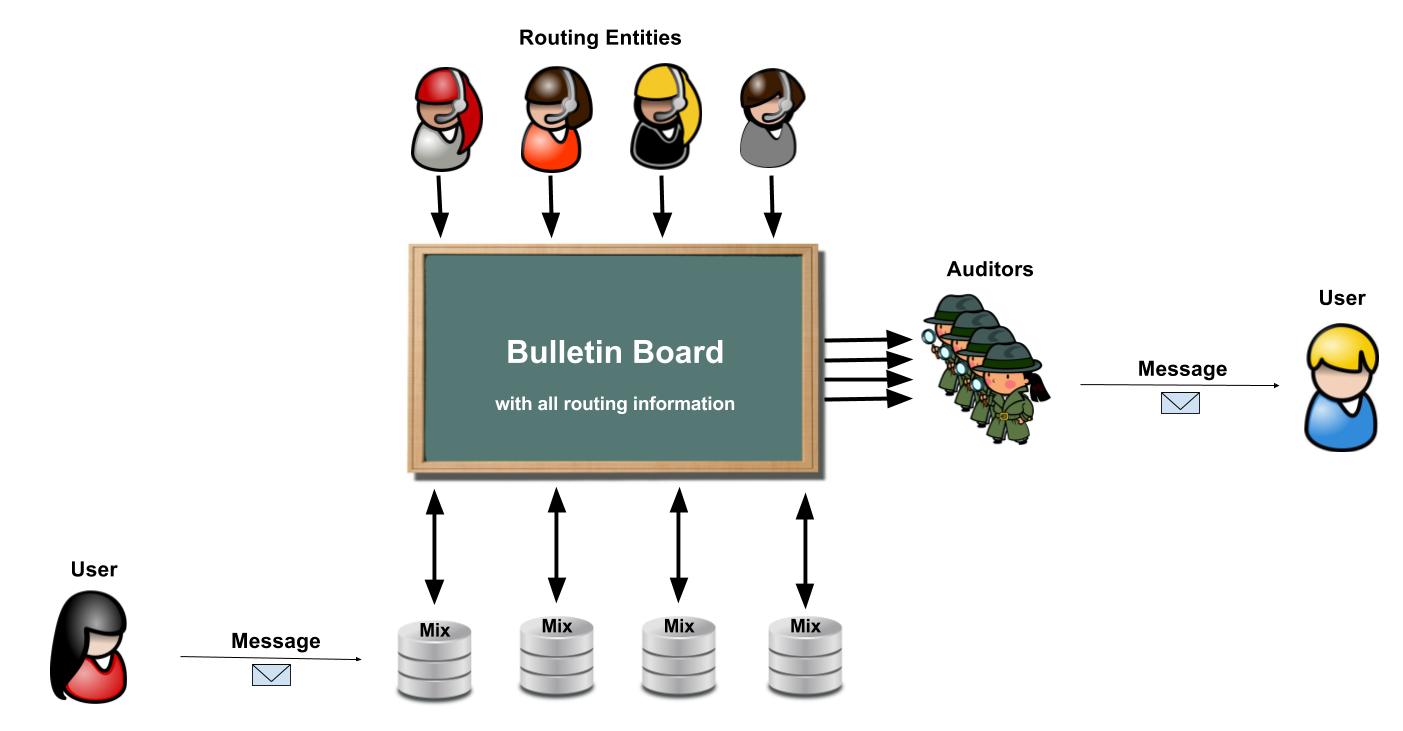}
\caption{This figure depicts the roles of all actors (user, mixes, ruting entities, auduting servers) as a message is relayed through the system. }
\label{fig:System}
\end{figure}

Our system comprises a set of users and dedicated servers as depicted in Fig.~\ref{fig:System}. The users in the system want to send messages anonymously, whereas the servers cooperate with each other to achieve secure anonymous routing.  
We distinguish between three sets of servers: $\mathcal{R}$ - relays,  $\mathcal{RE}$ - routing entities, and $\mathcal{AS}$- auditing servers.  The relays $\mathcal{R}$ are mixes that relay messages via publishing ciphertext messages on the bulletin board and fetching them. More precisely, the mixes first collect from the bulletin board their set of input ciphertexts, then re-encrypt and shuffle those, a process referred to as \emph{mixing}, and then relay their outputs to the bulletin board, so that each of them can be retrieved by the next mix in the path. The relays are organized in $l$ sequential layers formed by disjoint subsets of $\mathcal{R}$.

The second set of servers are called \emph{routing entities} $\mathcal{RE}$. These are responsible for computing the joint randomness that is used in the routing process.

The third set of servers are the \emph{auditors (auditing servers)} $\mathcal{AS}$. Auditors are responsible for: 1. generating and broadcasting the public encryption key; 2. verifying the correctness of message shuffling and routing and 3. final message decryption.

For clarity, we keep the roles of auditors and routing entities separate from mixes. However, from a practical point of view all three roles can be carried out by the same set of severs.


Next, we briefly review the required cryptographic building blocks and proceed by describing our protocol. 

\subsection{Cryptographic Building Blocks}
Next, we review relevant cryptographic building blocks.

\textbf{Public Key Encryption and Signing} 
 We assume the reader is familiar with the principles of pubic key encryption and signature schemes.
In our protocol, we employ a variant of public key encryption, referred to as \emph{threshold decryption} \cite{Desmedt:1989:TC,Shamir:1979:SS}, which allows for the decryption to be carried out by a quorum of participants (the auditors in our scheme).   
More precisely, our scheme uses a distributed key generation protocol \texttt{KeyGen}, such as the ones of \cite{Pedersen:1991:DKG,Gennaro:1999:DistKeyGen}. It generates a public key which is made available to all users of the system, while the corresponding secret key is distributed to all the auditors $\mathcal{AS}$ in a secret shared form.
During the decryption protocol \texttt{Dec} a threshold of the secret key share holders (auditors) jointly decrypt the input ciphertext. This process can be  can be realized either with a  trusted dealer, such as  in the work of \cite{Shoup:1998:AnalysisTDH}, or without \cite{Pedersen:1991:TCWithoutDealer}, \cite{Cramer:1997:DistributedCrypto}. 


To conceal how ciphertexts traverse a mix, we use re-randomizable encryption, such as ElGamal \cite{ElGamal:1985}. This allows a mix to update the randomness of the traversing ciphertexts. We propose standard cryptographic signatures, such as \cite{Schnorr:1989:Signatures}, for the routing verification.

\textbf{Shuffle and Correctness of Shuffling} A shuffle algorithm \texttt{Shuffle} implements a permutation on its inputs. In our protocol we use shuffles to permute the ciphertexts to conceal their ordering.
To guarantee the correct execution of the shuffling algorithm we apply a \emph{proof of correct shuffling} which is realized by a zero-knowledge (ZK) proof systems.
In our scheme, given a set of pre-shuffling and post-shuffling ciphertexts, a prover (the mix) needs to prove to a verifier (the auditors or, in case the auditor role is carried out by mixes, other mixes) that the set of post-shuffling ciphertexts is a re-randomization of the set of the pre-shuffling ciphertexts (without revealing the permutation and the re-randomization randomness used by \texttt{Shuffle}).
Our protocol utilizes non-interactive variants of ZK proofs, such as the ones proposed in \cite{Wikstrom:2009:CommitmentProofofShuffle}\cite{Bayer:2012:EfficentZKProofofShuffle}.


\section{Our Multiparty Routing Protocol}\label{sec:protocol}

\subsection{System Overview}
As mentioned before, mixes are divided into a number of subsets (layers).
Messages are relayed by one mix from each layer of the system. Therefore, each message is relayed by a set of mixes.
Users send their messages to an arbitrary mix from the first layer of mixes $\mathcal{R}_1$. 
Only messages that are submitted within the same time frame are mixed with each other.
Hence, a user is only anonymous among honest users that submitted their messages at the same time frame. 

Each relay mixes all message it receives.
The next mix is determined in a hop-by-hop fashion.
However, the mixes do not have freedom to choose to which mixes they send any given output message.
The routing entities, $\mathcal{RE}$, determine how each message is routed by jointly producing the randomness used for the routing; this process, called \emph{Multiparty Routing}, can be summarized as follows.
\begin{enumerate}
 \item All routing entities choose privately a random number and use a
   commitment scheme to commit to their random numbers on the bulletin board.

 \item After all routing entities have committed to their random numbers, they all open their commitments.

 \item The random numbers are combined to produce a joint random number.

 \item This random number is used to compute the routing assignment of the ciphertext messages to the mixes in the next layer in proportion to their capacity/throughput.

 \end{enumerate}
This procedure is carried out for the output of each mix. In parallel to this,
the auditing servers need to verify the correctness of the shuffling and routing.
After a message has been mixed by a mix from each layer and the auditing servers successfully verify the correctness of all routing and shuffling processes the message is delivered to the intended recipient of the message.
If at least one of the mixes that has relayed the message is honest, the adversary cannot trace back a message leaving the network to any particular sender.

\begin{figure}[h]
    \centering
\includegraphics[width=0.35\textwidth]{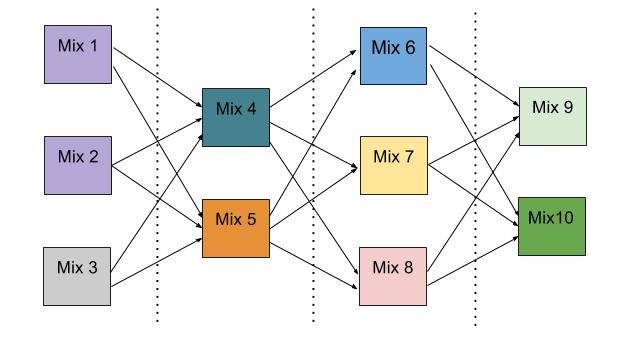}
\caption{In this figure we present an example of how mixes can be arranged for our protocol when the protocol consists of $4$ layers. The throuput of all layers needs to be equal, meaning that the total throughput of mixes $1,2,3$ needs to be equal to the total throuput of the mixes $4,5$, and the total throuput of the mixes $6,7,8$, ad the total throuput of the mixes $9,10$. For example, the output ciphertexts from mix $1$ are either assigned to mixes $4$ or mix $5$. }
\label{fig:Topology}
\end{figure}
Figure \ref{fig:Topology} shows the topology for a network that has four layers (disjoint subset) of mixes.
After processing input messages, the MPR protocol assigns output messages to mixes in the next layer;
meaning that a portion of the output messages of each mix is assigned to mixes in the next layer in proportion to the throughput of those mixes within their own layer.
Hence, the routing pattern between the mixes follows a stratified topology as investigated by Diaz et al. \cite{Diaz:2010:TopologyComparisonMixnets}.

\subsection{Main Protocol}

We denote the set of users by $U$.
The set $\mathcal{R}$ of relays is divided into $l$ layers formed by mutually exclusive subsets $\mathcal{R}_i$ with the union $\bigcup_{i=1}^{l}{\mathcal{R}_i} = \mathcal{R}$. An individual mix is denoted by $r$.

The bulletin board contains lists of routing and verification information (input and output
ciphertexts, corresponding signatures, commitments and their
corresponding openings), and NIZK proofs of shuffles, grouped and indexed by $ctr$ going from $1$ to $l$).

\paragraph*{Encryption} The auditors (auditing servers) $\mathcal{AS}$ are responsible for the generation of the encryption key $pke$.
Each auditing server has only a share of the private key $\{skd_{\mathcal{AS}_{1}},...,skd_{\mathcal{AS}_{d}}\}$ where $|\mathcal{AS}| = d$.  
Only qualified subsets of the auditing servers can jointly decrypt ciphertexts using their private key shares.

The protocol starts with each user encrypting her message $m$ together with its intended receiver under the public key $pke$ (known to all users) and sending the corresponding ciphertext to an arbitrary mix of $\mathcal{R}_1$.
The set of ciphertexts that enter the system in the same time (by all senders) is denoted by $C_0$.
Our system mixes ciphertexts in multiple layers with layer $\mathcal{R}_i$ producing the ciphertext set $C_i$.
After completing $l$ layers the protocol terminates with a verification step and the decryption of $C_l$.

\paragraph*{Mixing} Once a subset of the ciphertexts are assigned to a mix $r$ the \emph{mixing} and \emph{routing} take place.
Mixes in the first layer receive ciphertexts from users, while mixes for other layers take ciphertexts from the bulletin board.
The mixing consists of the \emph{re-encryption} of the ciphertext inputs and their \emph{shuffling} by a random permutation $\phi$.

After the output ciphertexts have been submitted to the bulletin board, the next
mix to shuffle a ciphertext is determined by the following \emph{routing assignment} procedure.

\paragraph*{Routing Assignment} We use a coin-tossing protocol such as
\cite{Blum:1983:CoinFlipping} to  obtain joint randomness.
In the coin-tossing protocol the routing entities commit to their randomness
using a cryptographic commitment schemes. Once the commitments are opened and the random strings are revealed on the
bulletin board, a simple function, e.g. bitwise XOR or modular addition, is applied to obtain the joint randomness $Rand$.

Then we apply a function $\texttt{Assign}$ on $Rand$. $\texttt{Assign}$ then outputs the next layer mix index assignment list $\mathcal{L} =(z_1, \ldots, z_w)$ with $k$ elements in correspondence with the capacity/thoughput of these mixes.

We give an example instantiation of $\texttt{Assign}$ via the two procedures $\texttt{H}$ and $\texttt{Map}$ in Appendix A in Algorithms \ref{algorithm:hashing} and \ref{algorithm:mapping}.

\begin{figure}[h]
    \centering
\includegraphics[width=0.35\textwidth]{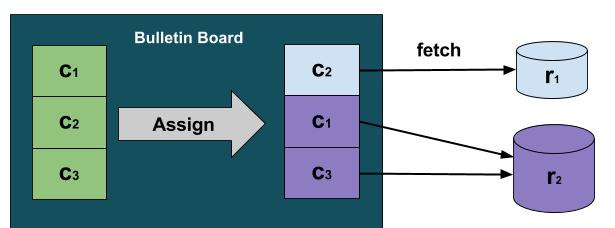}
\caption{This figure depicts how the ciphertexts $c_1, c_2, c_3$ from our example are assigned to the mixes $r_1$ and $r_2$.}
\label{fig:mapping}
\end{figure}

The example in Figure \ref{fig:mapping} to illustrates the routing assignment. Let us have a relay which outputs ciphertexts $c_1, c_2, c_3$ and the next layer of relays consist of two relays $r_1, r_2$ with corresponding capacities $b_1=1,b_2=2$.
By applying \texttt{Assign} we want to determine which relays in the layer receive which ciphertexts.
Namely, for  an output vector of indices ($z_1=2$, $z_2=1$, $z_3=2$) output by $\texttt{Assign}$ the outgoing ciphertext is going to be fetched in the next step as follows. In our case, $r_1$ will then fetch the first $\frac{1}{1+2}=\frac{1}{3}$ of the outgoing ciphertexts, namely $c_{z_1} = c_2$, and $r_2$ fetches the rest $\frac{2}{1+2} = \frac{2}{3}$ ciphertexts, $c_{z_2} =c_{1}$ and $c_{z_3} =c_{3}$ respectively.

\paragraph*{Proof of Mixing and Routing} The correctness of the mixing process is accommodated by the bulletin board. 
After mixing the set of input and corresponding set of output ciphertexts the mix needs to sign input and output ciphertexts. 
For signing, both the routing and mixing, $\mathcal{R}$ use their signing private key $sks$. The corresponding verifying public key $pkv$ is available to all parties for later verification purposes. 
These keys are used as long-term keys, unlike $(pke, (skd_1,\ldots,skd_d))$ that are issued fresh for each time-frame. 

Then, to prove correctness of protocol execution each mix needs to post on the bulletin board: a) inputs and outputs and their corresponding signatures, and b) a non-interactive zero-knowledge (NIZK) proof of correct mixing.
Also, the bulletin board maintains the commitments for the randomness used for the MPR routing algorithm and their subsequent openings posted by the routing entities.

\paragraph*{Verification and Decryption}



Ciphertext which have passed through all layers are decrypted by the auditing servers upon successful verification of the correctness of: 1) Mixing and 2) Routing.

After all mixes in a layer published their results on the bulletin board, the auditing servers verify their mixing process by verifying all NIZK proofs on the bulletin board.
Moreover, the auditing servers carry out routing verification on the input/output ciphertext signed on the bulletin board for each layer as soon as input ciphertexts for the next layer start being published to the bulletin board.
The routing verification takes the randomness (commitments openings) and computes the assignment for the next mix layer.
Then, correct routing is confirmed by verifying that the computed assignment corresponds to the correct input and output ciphertexts in the bulletin board.
With Algorithm  \ref{algorithm:verifyrouting} we give an example instantiation of the routing verification algorithm in Appendix B.

If routing or mixing verification fails, meaning that a mix has taken wrong ciphertext as input or produced an invalid NIZK proof, the auditing servers record the verification failure of the corresponding mix on the bulletin board.
If routing and mixing of a mix does not verify the routing entities generate a fresh $Rand$ to assign the input ciphertexts from the mix that has performed wrong routing to the remaining mixes in his layer.
All remaining output ciphertexts in the layer need to wait until this stage is carried out to be routed to next layers.

The integrity and authenticity of bulletin board submissions is verified by the use of the signatures schemes.
In our protocol we use randomized algorithms and with $\getsr$ we denote the
randomized output generation, so $\getsr S_w$ samples randomly from all permutations
of size $w$ and $s$ refers to the randomness used by the ciphertext re-randomization algorithms. 

Our protocol is summarized in Algorithm~\ref{alg:protocol}.
\begin{algorithm}
  \small
\caption{Multiparty Routing}\label{algorithm:MPR}
\text{ \noindent \hspace{1ex}} \textbf{\emph{Setup}}\newline
\text{ \noindent \hspace{1ex}} Auditing servers $\mathcal{AS}$ generate encryption keys: \newline
\text{ \noindent \hspace{3ex}} $(pke,skd_{{\mathcal{AS}_{1}}},...,skd_{\mathcal{AS}_{d}}) \getsr$ \texttt{KeyGen} \newline
\text{ \noindent \hspace{3ex}} publish $(pke)$ to bulletin board \newline

\text{ \noindent \hspace{1ex}} \textbf{\emph{Encrypt}}\newline
\text{ \noindent \hspace{1ex}} Each sender: \newline
\text{ \noindent \hspace{3ex}} $c \getsr$ \texttt{Enc}$_{pke}($receiver's address$, m)$ \newline
\text{ \noindent \hspace{3ex}} sends $c$ to $r \in \mathcal{R}_1$ \newline
\text{ \noindent \hspace{3ex}} $ctr \gets 1$ \newline

\text{ \noindent \hspace{1ex}} \textbf{while} $ctr\leq l$ do \newline

\text{ \noindent \hspace{3ex}} \textbf{\emph{Mixing and Prove Mixing}}\newline
\text{ \noindent \hspace{5ex}} Each $r \in \mathcal{R}_{ctr}$, with inputs $C = \{c_1,
\ldots, c_w\}$ and $\phi \getsr S_w$: \newline
\text{ \noindent \hspace{7ex}} shuffles $C^\phi \getsr$ \texttt{Shuffle}$(C, \phi)$ \newline
\text{ \noindent \hspace{7ex}} re-encrypts  $(C', s)\getsr$ \texttt{ReEnc}$(C^\phi)$ \newline

\text{ \noindent \hspace{5ex}} Each $r \in \mathcal{R}_{ctr}$:\newline
\text{ \noindent \hspace{7ex}} computes $\pi \getsr$ \texttt{Prove}$_{\text{NIZK}}(C, C'; \phi, s)$ \newline
\text{ \noindent \hspace{7ex}} publish $\pi$ to bulletin board \newline
\text{ \noindent \hspace{7ex}} publish (\texttt{Sign}$_{skv_{r}}(C),$ \texttt{Sign}$_{skv_{r}}(C'), C,C')$ \newline
\text{ \noindent \hspace{7ex}} to the bulletin board \newline

\text{ \noindent \hspace{3ex}} \textbf{\emph{Next hop and Prove Routing}}\newline
\text{ \noindent \hspace{5ex}} For each $r \in \mathcal{R}_{ctr}$, all $v$ routing entities $\mathcal{RE}$: \newline
\text{ \noindent \hspace{7ex}} publish \texttt{Com}$(rand_1),\ldots,$ \texttt{Com}$(rand_v)$ to the bulletin board \newline
\text{ \noindent \hspace{7ex}} publish $rand_1, \ldots , rand_v$ to the bulletin board

\text{ \noindent \hspace{5ex}} For each $r_{out} \in \mathcal{R}_{ctr}$, each $r \in \mathcal{R}_{ctr+1}$ computes: \newline
\text{ \noindent \hspace{7ex}} $Rand\gets$ \texttt{SUM}$(rand_{1}, \ldots, rand_v)$ \newline
\text{ \noindent \hspace{7ex}} $(z_1,\ldots,z_w)\gets$\texttt{Assign}$(Rand)$ \newline
\text{ \noindent \hspace{7ex}} For $i = 1$ to $w$  $r_{z_i} \in  R_{ctr+1}$ fetch $c_{i}$ \newline \newline
\text{ \noindent \hspace{3ex}}\textbf{\emph{Verify Mixing and Routing}}\newline
\text{ \noindent \hspace{5ex}} For each $r \in \mathcal{R}_{ctr}$, a threshold $z$ of auditing servers $\mathcal{AS}$ runs \newline
\text{ \noindent \hspace{7ex}} \texttt{Verify}$_{NIZK}(C_r, C'_r, \pi_r)$ \newline
\text{ \noindent \hspace{7ex}} \texttt{Verify}$_{pkv_{r}}$(\texttt{Sign}$_{sks_{r}}$ ($C_r$),$C_r$) \newline
\text{ \noindent \hspace{7ex}} \texttt{Verify}$_{pkv_{r}}$(\texttt{Sign}$_{sks_{r}}$ ($C'_r)$,$C'_r$) \newline
\text{ \noindent \hspace{5ex}} and computes \newline
\text{ \noindent \hspace{7ex}} \texttt{VerifyRouting}$(Rand_r,C'_r, \mathcal{R}_{ctr+1})$ \newline \newline
\text{ \noindent \hspace{3ex}} $ctr\gets ctr + 1$ \newline
\text{ \noindent \hspace{1ex}} \textbf{end while} \newline

\text{ \noindent \hspace{1ex}} \textbf{\emph{Decrypt}}\newline
\text{ \noindent \hspace{3ex}} Each $r \in \mathcal{R}_l$: \newline
\text{ \noindent \hspace{5ex}} sends all $c'$ to $\mathcal{AS}$ \newline
\text{ \noindent \hspace{3ex}} For a threshold $z$ of auditing servers $\mathcal{AS}$: \newline
\text{ \noindent \hspace{5ex}} if \emph{\textbf{Verify Mixing and Routing}} is successful for all layers \newline
\text{ \noindent \hspace{7ex}} decrypts $($receiver's address$,m) \gets$ \texttt{Dec}$_{skd_1, \ldots skd_{z}}(c')$ \newline
\text{ \noindent \hspace{5ex}} otherwise returns an error\newline
\text{ \noindent \hspace{5ex}} send $m$ to $receiver$ from $U$ \newline
\label{alg:protocol}
 \end{algorithm}



\subsection{Parameter Choices}
Our protocol has similarities to parallel mixing in terms of distributing ciphertexts among a set of mixes, hence, we revisit suggested parameters by the literature and adopt these parameters for our system.



The number of mixes that relay the ciphertexts, or so-called path length, is an important aspect of our system because it determines how many times a batch of ciphertexts $C$ is re-arranged. The path length has a direct effect on whether a ciphertext batch $C$ entering the system  is fully mixed with the other batches entering at the same time. Our result confirms the intuition that a  larger number of layers leads to a better mixing.

For a passive adversary, Czumaj shows that the path length in any switching networks should be $\log n$, where $n$ is the total number of mixes in the system~\cite{Czumaj:2001:SwitchingNetowrksRandomPermutations} \cite{Czumaj:2015:RandomPermutationSwitchingNetworks}.
Similarly, Klonowski et al.~\cite{Klonowski:2005:ProvableMixnets} have also shown that in parallel mixing the number of distribution steps should be $\log n$.

For an active adversary, Klonowski~\cite{Klonowski:2005:ProvableMixnets} shows that the number of mixing layers must 
depend on the number of messages that are relayed through the system.
Similarly, Goodrich and Mitzenmacher~\cite{Goodrich:2012:AnonymousCardShuffling} have shown that if some mixes are controlled by an active adversary then the number of layers needs to be $\log N$ where $N$ is the number of input messages in a given time frame.

Since, we are considering an active adversary we set the number of layers to $\log N$. For example, if we assume the routing of 1000 messages our system needs at least 3 layers where each layer has the throughput to relay 1000 messages. For relaying 100,000 messages the system needs to have at least 5 layers where each layer has the throughput to relay 100,000.messages.
To improve availability each layer should have at least two mixes.

Furthermore,  to increase the cost of an adversary  we assume that mixes are set up across distinct IP subnets by multiple organizations or institutions.
For example, an organization can offer more than one mix for the network, however, all mixes of an organization need to be in the same layer.
This is important because otherwise a message can be effectively mixed by a single organization.

Another important parameter is the capacity or throughput of the mixes.
From a load balancing point of view the total throughput of mixes in each layer needs to be equal in order to prevent any congestion in our system.


\section{Security Analysis}
 In this section we start by revisiting the security properties of interest and go on to analyze the security of our protocol based on those security goals against a global, active adversary. We show that an active adversary in our solution has no advantage over a passive one. 

\paragraph*{Security Properties}
\begin{itemize}
 \item \emph{Routing Integrity}: guarantees that an active adversary is not able to influence the routing decision to deanonymize messages by forcing them to route through adversarial mixes.
 \item \emph{Routing Correctness}: guarantees the detection of malicious routing and subsequently only correctly routed messages leaving the system. 
 \item\emph{ Anonymity} \emph{(routing confidentiality)}: guarantees that the adversary cannot find the relationship between messages leaving the system and a user sending a message to the system, provided that the message has traversed at least some honest mixes.  
 \item \emph{Availability}: guarantees that the protocol is robust against removal of a subset of entities by an adversary and that the messages entering the system will be output by the system in the face of such attacks.
 \end{itemize}


\subsection{Routing Integrity}
To address the integrity property we answer the following question:
 \emph{Can a malicious mix (or a subset of mixes) influence the routing of a ciphertext $c$ to a mix of his choice $r_{\mathcal{A}}$?}
We investigate two ways of manipulating the routing decision. The first way deals with adversaries who try to influence the shuffling procedure. This attack path is however not possible in our protocol because all mixes publish their shuffled ciphertexts before the routing decision ($Rand$ outcome) is revealed. The second way to manipulate the routing decision is through adversarial bias in the computation of the joint randomness $Rand$.


More concretely, we investigate the advantage of an active adversary $\mathcal{A}$ who gains control over a subset of routing entities denoted as $\mathcal{RE}_A$. In order to influence the routing decision $\mathcal{A}$ has to be able to bias the computation of $Rand$ where $Rand = Rand_{honest} + rand_{\mathcal{A}}$ is the mod sum addition of the random numbers from all honest entities $Rand_{honest}$  and the adversarially controlled randomness $rand_{\mathcal{A}}$. The goal of $\mathcal{A}$ is to produce a new $rand'_\mathcal{A}$, such that the updated sum $Rand' = Rand_{honest} + rand'_{\mathcal{A}}$ fits its routing purposes: the target ciphertext is routed to $r_{\mathcal{A}}$ where $\mathcal{A}$'s capacity/throughput is $b/B$.
In our protocol \emph{all} routing entities $\mathcal{RE}$ first need to \emph{commit} to their random numbers and only then \emph{reveal }the openings of their commitments. $\mathcal{A}$ has control only over its own commitment $T = \texttt{Com}(rand'_{\mathcal{A}})$ and the corresponding opening $rand'_{\mathcal{A}}$. The best strategy for $\mathcal{A}$ is to wait for the openings of all honest routing entities and then reveal his own $rand'_{\mathcal{A}}$.

Since $\mathcal{A}$ does not know in advance the openings of the honest routing entities, $\mathcal{A}$ needs to `predict' in advance a valid $rand'_{\mathcal{A}}$, such that $T = \texttt{Com} (rand'_{\mathcal{A}})$).
Our assumption here is that $\texttt{Com}$ is a secure scheme against collision, preimage and second preimage attacks. Under this assumption $\mathcal{A}$'s best strategy is to perform precomputation to find the largest set of preimages $\mathcal{N}$ with $|\mathcal{N}| = n$ mapping to a commitment value $T$.  Then, once $\mathcal{A}$ commits to $T$, he `hopes' that the needed randomness for his routing purposes $rand'_{\mathcal{A}}$ belongs to the precomputed set $\mathcal{N}$.
Then, what is left to find is the size of the set of valid randomness values $\mathcal{V}$ for the routing purposes of $\mathcal{A}$. Since the adversary controls $b/B$ of the throughput $w$ ciphertexts, then the set of valid values $\mathcal{V}$ amounts to  $\frac{2^{|Rand|}}{w}\times \frac{b}{B}\times w$ where $\frac{2^{|Rand|}}{w}$ is the fraction of random values resulting in a given router assignment as defined by our routing algorithm.
Finally, the probability that $\mathcal{A}$'s set of precomputed values $\mathcal{N}$ ends up intersecting with one of the valid random values $rand'_{\mathcal{A}}$ is $\frac{n}{2^{|Rand|}\times b/B}$. In other words, $\frac{n}{2^{|Rand|}\times b/B}$ amounts to the probability that  $\texttt{Com}^{-1}(T) = rand'_{\mathcal{A}}$ where $rand'_{\mathcal{A}} = (Rand' - Rand_{honest}) \in \mathcal{V} \cap \mathcal{N}$.

Note that, as we assumed, if $\texttt{Com}$ is collision secure, finding $n = 2$ values takes already precomputation time of birthday bound complexity $2^{|Rand|/2}$. This means that unless $\mathcal{A}$ has a precomputational power of order $2^{|Rand|/2}$  its success probability in biasing the joint randomness $Rand$ is  $\frac{1}{2^{|Rand|}\times b/B}$.




Furthermore, if the adversary removes the mix that $c$ is assigned to to force a new routing assignment his chances to be assigned to $c$ are increasing only if he removes a large proportion of throughput of the layer of $r_{\mathcal{A}}$.

\subsection{Routing Correctness}

In this scenario we investigate if a mix can unnoticeably  deviate from the routing assignment.
The adversary goal (acting as one of the mixes $\mathcal{R}$) is to get a ciphertext of  his choice regardless of the ciphertext's next hop  assignment. To achieve her goal the adversary here needs to manipulate the \emph{routing verification}. To do so the adversary needs to control a subset of the auditing entities that read the routing information from the bulletin board and carry out the verification. But this is impossible in our protocol because this adversarial behavior is going to be discovered in the routing verification phase if at least one of the auditing servers is honest. Remember that one of the tasks of the auditing servers is the verification of the correct input/output ciphertext routing and their signatures.


\subsection{Anonymity}
The adversary's aim is to trace the path of a ciphertext $c$ until it leaves the system.
To link an input ciphertext to an output ciphertext the adversary needs to control all mixes that have relayed the input ciphertext. Otherwise, if at least one mix on the ciphertext's route is honest, the adversary will not be able to trace this link.
Hence, our adversary needs to control at least one mix in each layer of the system.
Our decision to follow a stratified topology disallows the adversary to gain any advantage from placing his efforts (i.e. throughput) unevenly as compared to balancing out its efforts for each layer.  For example, if an adversary controls 20 $\%$ of the total throughput of the system, we assume that he is in fact controlling a 20 $\%$ of the throughput of the mixes in each layer.

To estimate the probability that a ciphertext follows a fixed path, denoted by $c\rightarrow p$, then we just need to multiply the probability of that ciphertext being routed by the individual mixes in each layer of the path.
Let us assume that a path $p$ consists of an order list of mixes $p=\{r_1,r_2, \ldots, r_l\}$, where $l$ is the number of layers of the system.
$$Pr[c \rightarrow p]=\prod_{i=1}^{l} \frac{b_{i}}{B}.$$ 
Moreover, the probability of choosing each of the mixes in a path is computed as follows.
where $b_i$ is the relative throughput of the mix $r$ compared to the total throughput $B$ of all mixes (recall that it is the same for all layers).
If all mixes in the subset have the same capacity $b$, then $Pr[c \rightarrow p]= (\frac{b}{B})^l$

\begin{figure}[h]
    \centering
\includegraphics[width=0.35\textwidth]{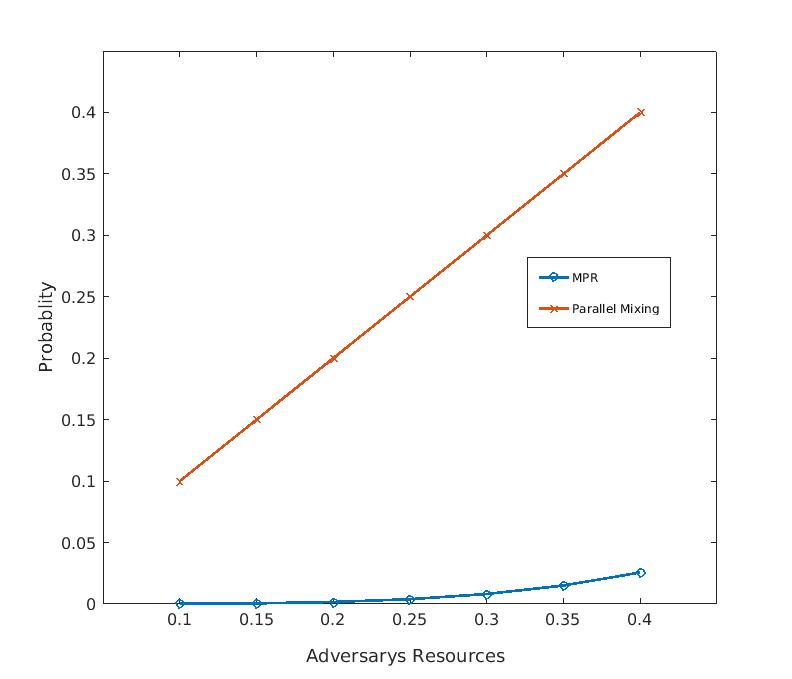}
\caption{This figure depicts the probability of a message being routed through mixes that are all controlled by the adversary and shows that this probability increases for stronger adversaries in MPR and in parallel mixing. For both MPR and regular parallel mixing the system has 4 layers. The adversary is controlling mixes in each layer by a proportion of 0.1, 0.15, 0.20, 0.25, 0.30, 0.35, 0.40.}
\label{fig:differentadversaries}
\end{figure}

We are interested in the probability that a ciphertext is routed only by adversarial mixes because this leads to deanonymize the ciphertext leaving the system by the adversary.
We compare the probability that the ciphertext relayed by our MPR protocol is only routed through adversarial mixes to to the probability that the ciphertext relayed by regular parallel mixing system is only relayed by adversarial mixes. 
Figure \ref{fig:differentadversaries} shows the probability that a ciphertext routed through only adversarial mixes, for our MPR protocol and for parallel mixes system, when the adversary is controlling $0.1 \%, 0.15 \%, 0.20 \%, 0.25 \%, 0.30 \%, 0.35 \%,$ and $0.40 \%$ of the throughput in each layer of the system. For this figure we assume the system has 4 layers. 
When regular parallel mixing is used the probability that a ciphertext is only routed along adversarial mixes is in proportion to the resources of the adversary. 
While if our MPR protocol is used, this probability is increasing  much slower due to the fact that adversarial mixes cannot steer ciphertexts toward further adversarial mixes. 


\begin{figure}[h]
    \centering
\includegraphics[width=0.35\textwidth]{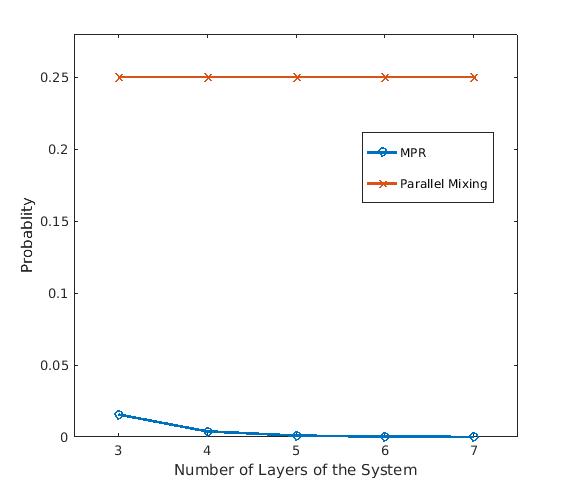}
\caption{This figure depicts the probability of a message being routed through mixes that are all controlled by the adversary and shows that this probability increases when the system has more layers of mixing. In both MPR and regular parallel mixing, the adversary is controlling 0.25 of mixes of each layer.}
\label{fig:differentnumberoflayers}
\end{figure}

Figure \ref{fig:differentnumberoflayers} shows the probability that a ciphertext is routed through only adversarial mixes, for our MPR protocol and for regular parallel mixes system, when the system has $3, 4, 5, 6,$ or $7$ layers. 
We assume for this figure that the adversary is controlling $0.25 \%$ of the throughput of each layer of the system. 
If regular parallel mixing is used, the probability that a ciphertext is only routed along adversarial mixes is constant and equal to the proportion of adversarial mixes irrespective of the number of layers. 
Wile when our MPR protocol is used, this probability is significantly lower compared to parallel mixes even when the system consists of only 3 layers. This probability decreases further  as the number of layers in the system increases. 

Moreover, recall that in Parallel Mixes it is required to distribute the input messages uniformly before submitting them to the system to decrease the chances of carrying out attacks such as blending attacks.   
However, permuting incoming messages before entering the system to obtain a uniform distribution would need another randomly distributing mechanism and complicates the trust model.
Furthermore, Parallel Mixes require two rotating phases before and after a single re-distribution phase where the input ciphertexts are rotated $m+1$ times, where $m$ is the number of malicious mixes.
Using MPR there is no need to enter only uniformly distributed messages to the system and the rotating phases of Parallel Mixes can be eliminated;
this allows to use multiple distribution phases instead of only one, which contributes to obtaining a complete random permutation of all ciphertexts.

\subsection{Availability}
The goal of the adversary is to take out stakeholders of the protocol in order to obstruct the protocol from running.

There is a restriction on how many mixes, routing entities, and auditors the adversary can take out.
Taking out auditors, $\mathcal{AS}$ is least beneficial, because the attack is not resulting in any advantage for the adversary.
While taking out routing entities, $\mathcal{RE}$ increases the adversary's success in biasing routing to a very small extent.
Taking out mixes will not abrupt the protocol execution but leads to traffic congestion.
Below we investigate the effect of removing each type of entity from the system.

\emph{Routing entities $\mathcal{RE}$}: If the routing entity fails before committing to a random number the protocol continues to operate as long as it is assumed that at least one honest routing entity has remained.
If the routing entity fails after committing to a random number it is considered an abortion, although it might have been a system failure.
If this happens more than once, the remaining routing entities vote the corresponding routing out.

\emph{Auditors $\mathcal{AS}$}: If a limited number of auditors fail to carry out the distributed decryption, then the protocol is guaranteed to continue execution because we are using threshold cryptography.
To carry out the routing and mixing verification the only restriction is that at least one remaining auditor is not controlled by the adversary.

\emph{Mixes $\mathcal{R}$}: Our protocol is resilient to the failure of a number of mixes and preserves load balanced among the remaining mixes.
We describe how the protocol recovers after the failure of a mix or a number of mixes in different positions as follows.
\begin{enumerate}
\item If a mix fails at the first layer after the users have sent their ciphertexts the users machine needs to notice in the bulletin board that her ciphertext is not in the input of any of the entry mixes and send their ciphertext to another entry mix. 
\item If the adversary takes out all mixes in the first layer the user's machine will notice that the ciphertext sent by her is not in the bulletin board and the auditors notice the absence of any output from the first layer and introduce new entry points for the users.  
\item If a mix fails at one of the layers except of the first layer before generating the output ciphertext and the corresponding routing information, the auditors notice this and inform the routing entities to generate a new $Rand$ that assigns the input ciphertext of the failed mix to the remaining mixes in its layer. 
\item If the adversary takes out all mixes in one layers except of the first layer, the auditors inform the next layer of mixes to take over with fresh random numbers. 
\end{enumerate}

\section{Discussion}

\textbf{Decentralization Advantages}: MPR decentralizes the routing decision which leads to several advantages over a routing decision that is not decentralized. 
Troncoso et al.~\cite{Troncoso:2017:Decentralization} present numerous advantages of decentralization in protocol design such as increasing the adversaries cost and facilitating public verifiability. 
Our MPR protocol also increases the cost of the adversary for attacking the routing integrity and facilitates public verifiability. 

 
 
\textbf{Latency} We estimate for a system with 9 mixes having equal throughput, arranged in 3 layers each layer having the throughput to relay 900 messages, and where 1000 messages were sent to the system, messages takes about 20 seconds to traverse the system. 
This is, however, only a preliminary result tested on a Intel Xeon E5-2640 CPU, 2.50 GHz machine with 24 cores, where all computation are performed online. 
If some of the computations are made offline the latency can improve further.  
Increasing the input traffic and consequently the number of layers of the system strengthens anonymity, but also increases the latency. There is a trade-off between latency and anonymity. Our system offers almost optimal load balancing, when congestion is minimized. As long as the adversary does not take out mixes the latency of the system thus increases linearly with the amount of traffic. 
Our protocol is a medium-latency anonymous communication network and thus can be deployed for applications that are more tolerant to latency such as anonymous wiki, micro-blogging, voting, and auctions. Note that for application such as voting with homomorphic tallying the decryption phase of our protocol may not be necessary for all messages.

\textbf{Load Balancing}: 
Since, our MPR protocol offers load balancing the messages leaving the system have a more uniform anonymity set size compared to mixnets using source routing because in source routing the routes are selected without considering load balancing. This increases the minimum anonymity that a message can achieve. 

\textbf{Limitation}
As mentioned before, MPR addresses the scalability issue of source routing in terms of not requiring to maintain a complete system view for users. 
Growing our system is, however, not as scalable as growing a system with a free topology because the system's throughput can be only expanded by adding mixes to all layers, while if there is no restrictions for selecting a relay and the routing can choose any relay in the system adding a single relay increases the system's overall throughput.

\section{conclusion}

We proposed a novel routing approach, called multiparty routing, to carry out secure routing for anonymous communication networks.
More concretely, our MPR protocol achieves routing integrity, routing correctness, load balancing, in addition to anonymity, and is relatively scalable.

We leave as future work a full implementation of our protocol and the addition of resilience mechanism against failing mixes explained in the security analysis.
Moreover, we leave as an open question measurements of the performance and resilience of our system under DoS attacks.

\begin{appendix}
\section{Appendices}
In this section, we give instantions of routing assignment and routing verification. 
\subsection{Routing Assignment}
To route a ciphertext, at each layer, a relay is assigned to the ciphertext using a mapping function. The mapping function is composed of a permutation function keyed by the jointly
generated random number $Rand$ and a load-balancing function that distributes ciphertext
based on the throughput of relays in the next layer.

The permutation algorithm is the function \texttt{H} out of hash function \texttt{h}.  
\texttt{H} is applied iteratively on the inputs: joint random number $Rand$ and the ciphertext indexes $1, \ldots, w$ and returns the permuted indexes $z_1, \ldots, z_w$. 
To realize the permutation functionality we have to avoid hash function collisions by performing a check of membership on a set $Z$ of already computed index values. 

\begin{algorithm} [t]
\caption{Permuting with Jointly Obtained $Rand$}\label{algorithm:hashperm}
\texttt{H}$(Rand, 1, ..., w)$ \newline
\text{ \noindent \hspace{1ex}} $Z=\emptyset$\newline
\text{ \noindent \hspace{1ex}} $i=1$ \newline
\text{ \noindent \hspace{1ex}} \textbf{for} $i=1$ to $w$ \newline
\text{ \noindent \hspace{3ex}} $j=0$ \newline
\text{ \noindent \hspace{3ex}} $z_{i}=h(Rand\|i\|j)$ \newline
\text{ \noindent \hspace{3ex}} \textbf{while} $z_{i}\in Z$ \newline
\text{ \noindent \hspace{5ex}} $j=j+1$ \newline
\text{ \noindent \hspace{5ex}} $z_{i}=h(Rand\|i\|j)$ \newline
\text{ \noindent \hspace{3ex}} \textbf{end while} \newline
\text{ \noindent \hspace{3ex}} $Z \leftarrow Z \cup z_{i}$ \newline
\text{ \noindent \hspace{1ex}} \textbf{end for} \newline
$output(z_1, \ldots, z_w)$ \newline
\label{algorithm:hashing}
\end{algorithm}

We apply the mapping function \texttt{Map}, e.g., Algorithm \ref{algorithm:mapping} in appendices, to make the assignment of the permuted ciphertext (indexed by \texttt{H}) to the relays in the next layer in proportion to their throughput. 
We refer to the throughput of relays by $b$, where throughput refers to the number of messages it can route in proportion to all the number of message that can be routed by a layer and the throughput of a single layer is denoted $B$. 
For this purpose we use the function \texttt{fetch}, which ciphertexts to mixes in the next layer. 
These relays fetch the ciphertext that is assigned to them as their input from the bulletin board.

\begin{algorithm} [t]
\caption{Mapping ciphertexts}\label{algorithm:mapping}
\texttt{Map}($z_1, ..., z_w$, $r_1, ..., r_k$, $b_1,...,b_k$) \newline
\text{ \noindent \hspace{1ex}} $B=$\texttt{SUM}$(b_1,...,b_k)$ \newline
\text{ \noindent \hspace{1ex}} \textbf{for} $j=1$ to $k$ \newline
\text{ \noindent \hspace{3ex}} $r_j$ \texttt{fetch} $w * bj/B$ ciphertexts from bulletin board \newline 
\text{ \noindent \hspace{1ex}} \textbf{end for} \newline
Assigns ciphertext to mixes in proportion to the mixes throughput.\newline
\label{algorithm:mapping}
\end{algorithm}

\subsection{Routing Verification}
Let $C_r$ refer to the set of input ciphertext that relay $r$ has submitted
and signed and $C'_r$ to the set of output ciphertexts that relay $r$ has
submitted and signed.
Moreover, let us denote the  $ctr$-th layer of routers in our system by $R_{ctr}$.

Then all the auditors must carry out the routing verification as described below. 
Our algorithm applies a Boolean \texttt{VerifyRand} function which simply
verifies both the correctness of the opening os the commitments \texttt{Com}$(rand)$ with regard to the
actual $rand$ on the bulletin board and checks that their sum equals $Rand$. 
Moreover, the algorithm verifies whether the input of all relays in the $ctr+1$ layer is correct according to the output ciphertexts of relays in layer $ctr$ and their corresponding $Rand$s. 
\begin{algorithm} [h]
\caption{Verify Routing}
\texttt{VerifyRouting}$(Rand_r$,$C'_{r}$, $R_{ctr+1})$ \newline
\text{ \noindent \hspace{1ex}} \textbf{if} \texttt{VerifyRand} returns $1$ \newline
\text{ \noindent \hspace{3ex}} $(c_1, \ldots, c_w)=\mid C'_{r}\mid$ \newline
\text{ \noindent \hspace{3ex}} $(z_1, \ldots, z_w)=$ \texttt{H}$(Rand_{r}, 1, \ldots, w)$ \newline
\text{ \noindent \hspace{3ex}} $(r_{1}, \ldots, r_{w})=$ \texttt{Map}$(c_1, \ldots, c_w, R_{ctr+1}, B_{ctr+1})$ \newline
\text{ \noindent \hspace{3ex}} \textbf{for} $j=1$ to $w$ \newline
\text{ \noindent \hspace{5ex}} \textbf{if} relay ${r_j}$ signed $c_j$ \newline
\text{ \noindent \hspace{7ex}} \textbf{continue} \newline
\text{ \noindent \hspace{5ex}} \textbf{else return} (``Routing error'', $r_j$) \newline
\text{ \noindent \hspace{5ex}} \textbf{end if} \newline
\text{ \noindent \hspace{3ex}} \textbf{end for} \newline
\text{ \noindent \hspace{1ex}} \textbf{else if} return (``Commitment error'')\newline
\text{ \noindent \hspace{1ex}} return (1) \newline
\label{algorithm:verifyrouting}
\end{algorithm}

\end{appendix}


\section*{Acknowledgment}

This work was supported in part by the Research Council KU Leuven: C16/15/058. In addition, this work was supported by the European Commission through KU Leuven BOF OT/13/070, H2020-DS-2014-653497 PANORAMIX and H2020-ICT-2014-644371 WITDOM. 
This work was supported by Microsoft Research through its PhD Scholarship Programme. 
This work was supported by the European Commission under the ICT programme with contract H2020-ICT-2014-1 644209 HEAT.


\bibliographystyle{ieeetr}


\bibliography{references}

\end{document}